\documentclass[%
 aip,
 amsmath,amssymb,
 reprint,%
]{revtex4-1}

\usepackage{graphicx}
\usepackage{dcolumn}
\usepackage{bm}
\usepackage{xcolor}

\usepackage[utf8]{inputenc}
\usepackage[T1]{fontenc}
\usepackage{mathptmx}

\newcommand{\prt}{\partial}

\newcommand{\la}{\lambda}

\newcommand{\om}{\omega}

\begin{document}

\title{Asymptotic integrability of nonlinear wave equations}

\author{A.~M.~Kamchatnov}
\affiliation{Institute of Spectroscopy, Russian Academy of Sciences, Troitsk, Moscow, 108840, Russia}
\affiliation{Skolkovo Institute of Science and Technology, Skolkovo, Moscow, 143026, Russia}

\begin{abstract}
We introduce the notion of asymptotic integrability into the theory of nonlinear wave 
equations. It means that the Hamiltonian structure of equations describing
propagation of high-frequency wave packets is preserved by hydrodynamic evolution of
the large-scale background wave, so that these equations have an additional integral of motion. 
This condition is expressed mathematically as a system of
equations for the carrier wave number as a function of the background variables.
We show that a solution of this system for a given dispersion relation of
linear waves is related with the quasiclassical limit of the Lax pair for the
completely integrable equation having the corresponding dispersionless and
linear dispersive behavior. We illustrate the theory by several examples.
\end{abstract}

\pacs{02.30.Ik, 05.45.Yv, 47.35.Fg}


\maketitle

\begin{quotation}
At first sight it seems that the physical explanation of solitonic propagation as a
manifestation of subtle balance of dispersive and nonlinear effects, on one side,
and the mathematical idea of complete integrability of nonlinear wave equations, on
the other side, are not related with each other. We show in this paper that a combined
consideration of two limiting situations---the dispersionless evolution of large-scale
waves and the propagation of high-frequency wave packets according to Hamilton's
optical-mechanical analogy---sheds new light on the complete integrability condition.
As a result, we formulate a much weaker condition of asymptotic integrability which
means that the Hamiltonian structure of equations for the packet's propagation is
preserved by the dispersionless evolution of the background wave providing thus an
additional integral of motion. If this criterion 
is fulfilled, then we arrive at the quasiclassical limit of the Lax pair for some 
completely integrable equation. This theory is illustrated by several examples.
\end{quotation}

\section{Introduction} 

As was explained by classics of physics in the 19{\it th} century
\cite{bouss-1872,rayleigh-1876,stokes-91,kdv-95}, the phenomenon of solitary wave
(or solitonic) propagation is a result of subtle balance of nonlinear and dispersive
effects of wave motion. Independently, the mathematicians of the 19{\it th} century
developed the idea of complete integrability of equations of Newton dynamics
(see, e.g., \cite{whitt-27} and references therein), and this idea was based on some
specific properties of Hamilton's formulation of classical mechanics. Apparently,
these two fundamental ideas had not been related with each other until the discovery of
the inverse scattering transform method (IST) of integration of nonlinear wave
equations \cite{ggkm-67,lax-68,zs-71}. Soon after this discovery, it was realized
\cite{zf-71,gardner-71} that the nonlinear wave equations solvable by the IST method
possess the property of complete integrability generalized to systems with infinite
number of degrees of freedom. Since then, the Hamiltonian approach to nonlinear wave
equations has become an important tool for their investigation and, as a result, the
soliton physics has become a well-developed part of the modern mathematical physics
(see, e.g., \cite{dickey,ft-07} and references therein). At the same time, it seems
that the original physical idea of interplay of nonlinear and dispersive effects has
not played any essential role in this formal development. The aim of this article is
to demonstrate that at least in some situations elaboration of this classical idea 
leads to interesting results useful for the nonlinear wave physics.

\section{Asymptotic integrability} 

We suppose that our physical system is described by
two variables which we will call for definiteness the ``density'' $\rho$ and the ``flow
velocity'' $u$. In one-dimensional geometry, they depend on the space $x$ and time $t$
variables: $\rho=\rho(x,t),u=u(x,t)$. In the dispersionless limit of long wavelengths,
the dynamics of our system is described by some hydrodynamic equations which can always
be transformed (in case of two dependent variables) to a diagonal Riemann form (see, e.g.,
\cite{ry-83})
\begin{equation}\label{eq2.1}
  \frac{\prt r_+}{\prt t}+v_+\frac{\prt r_+}{\prt x}=0,\quad
  \frac{\prt r_-}{\prt t}+v_-\frac{\prt r_-}{\prt x}=0
\end{equation}
for the so-called Riemann invariants $r_{\pm}=r_{\pm}(\rho,u)$. They are the most convenient
variables in wave physics, since in many typical situations the `velocities' $v_{\pm}$ are
given by the expressions
\begin{equation}\label{eq2.2}
  v_{\pm}=u\pm c,
\end{equation}
where $c$ is velocity of long sound waves propagating along a uniform background with
constant density $\rho$ and zero flow velocity $u$. This means that the velocities $v_{\pm}$
correspond to the upstream or downstream propagation of sound waves. Naturally, both $u$
and $c$ as well as $v_{\pm}$ can be expressed as functions of the Riemann invariants
$r_{\pm}$. Equations (\ref{eq2.1}) are nonlinear and they express the nonlinear properties
of our system.

The dispersive properties of the system are described by the dispersion relations
$\om=\om_{\pm}(k,r_+,r_-)$ (or $\om=\om_{\pm}(k,\rho,u)$) for harmonic linear waves
$\propto\exp[i(kx-\om t)]$ propagating along a uniform background with constant values
of $r_+,r_-$ (or $\rho,u$). It is important that in our formulation the dispersive and
nonlinear properties are not independent of each other: the velocities (\ref{eq2.2}) 
can be expressed as a long wavelength limit of the phase velocities of linear waves: 
\begin{equation}\label{eq2b}
  v_{\pm}=\lim_{k\to0}\frac{\om_{\pm}(k)}{k},
\end{equation}
so the two dispersion relations correspond to 
the two directions of propagation of linear waves.

Let now the background variables change little at distances of the order of the size
of the short-wavelength wave packet built from the linear waves with wave numbers around
the carrier wave number $k$. Then we can introduce the coordinate $x(t)$ of such a wave
packet and consider its motion along a nonuniform time-dependent background wave whose
evolution obeys Eqs.~(\ref{eq2.1}). In this case, the frequencies $\om_{\pm}$ become
slow functions of $x$ and $t$ via the background variables $r_{\pm}=r_{\pm}(x,t)$.
Consequently, propagation of our wave packet obeys the Hamilton equations (see, e.g.,
\cite{LL-2,ko-90})
\begin{equation}\label{eq3}
  \frac{dx}{dt}=\frac{\prt\om}{\prt k}, \quad \frac{dk}{dt}=-\frac{\prt\om}{\prt x},
\end{equation}
where $\om$ is given by one of the branches $\om_{\pm}$ of the dispersion relation.
Thus, to find the path of the wave packet, we have to solve at first the hydrodynamics
equations (\ref{eq2.1}), then substitute the solution into the dispersion relation
$\om(k,r_+,r_-)$, and, at last, to solve the Hamilton equations (\ref{eq3}).

The above general formulation of the packet's propagation problem refers to arbitrary
dispersion relations and related with them according to Eq.~(\ref{eq2b}) large-scale
hydrodynamic equations. Now we want to specify this formulation in such a way,
that it becomes `integrable'. This means that the Hamilton equations (\ref{eq3}) must have
an integral of motion which allows one to solve these equations in quadratures. Thus, we
demand that there exists such a function $k=k(r_+,r_-)$ that the Hamilton equations
(\ref{eq3}) are compatible with the background evolution according to the hydrodynamic
equations (\ref{eq2.1}). To express this condition in analytical form, let us consider
two moments of time separated by a small interval $dt$ and, consequently, by a small
distance $dx=v_gdt$ between the corresponding locations of the wave packet, where
$v_g=\prt\om/\prt k$ is the group velocity of the wave packet (see the first Eq.~(\ref{eq3})).
Then the differences of the values of the background Riemann invariants at the locations 
of the wave packet are equal to
$$
dr_{\pm}=\frac{\prt r_{\pm}}{\prt x}dx+\frac{\prt r_{\pm}}{\prt t}dt=
\frac{\prt r_{\pm}}{\prt x}(v_g-v_{\pm})dt,
$$
where we have taken into account the hydrodynamic equations (\ref{eq2.1}). This relation gives
$$
\frac{\prt r_{\pm}}{\prt x}dt=-\frac{dr_{\pm}}{v_{\pm}-v_g}
$$
and, consequently, the second Hamilton
equation (\ref{eq3}) yields the following expression for the small change of the wave number:
\begin{equation}\nonumber
  \begin{split}
  dk&=-\frac{\prt\om}{\prt x}dt=-\left(\frac{\prt\om}{\prt r_+}\frac{\prt r_+}{\prt x}+
  \frac{\prt\om}{\prt r_-}\frac{\prt r_-}{\prt x}\right)dt\\
  &=\frac{\prt\om/\prt r_+}{v_+-v_g}dr_++\frac{\prt\om/\prt r_-}{v_--v_g}dr_-.
  \end{split}
\end{equation}
On the other hand, our assumption that the wave number $k$ is only a function $k=k(r_+,r_-)$
of the background variables gives
$$
dk=\frac{\prt k}{\prt r_+}dr_+ +\frac{\prt k}{\prt r_-}dr_-.
$$
Since we consider an arbitrary general solution of Eqs.~(\ref{eq2.1}), the differentials $dr_+$
and $dr_-$ can be considered as independent of each other, so comparison of the above expressions
yields the system of equations
\begin{equation}\label{eq4}
  \frac{\prt k}{\prt r_+}=\frac{\prt\om/\prt r_+}{v_+-v_g},\quad
  \frac{\prt k}{\prt r_-}=\frac{\prt\om/\prt r_-}{v_--v_g},
\end{equation}
which define the function $k=k(r_+,r_-)$ of two independent variables $r_+,r_-$, which is
the desirable integral of motion. Naturally,
for existence of such a function these derivatives must commute,
\begin{equation}\label{eq5}
  \frac{\prt}{\prt r_-}\left(\frac{\prt k}{\prt r_+}\right)=
   \frac{\prt}{\prt r_+}\left(\frac{\prt k}{\prt r_-}\right),
\end{equation}
and this {\it asymptotic integrability condition} is only fulfilled for some special choices 
of the dispersion relation  $\om=\om(k,r_+,r_-)$ and, consequently, of the hydrodynamic 
equations (\ref{eq2.1}).

It is worth noticing that if we consider particular solutions of Eqs.~(\ref{eq2.1}) in the
form of simple waves, then one of the Riemann invariants is constant and the above
calculation leads to a single equation for the wave number $k=k(r)$ which becomes a
function of the only varying Riemann invariant $r$,
\begin{equation}\label{eq6}
  \frac{d k}{d r}=\frac{\prt\om/\prt r}{V-v_g},
\end{equation}
where $\om=\om(k,r)$ and $V=\lim_{k\to0}(\om(k)/k)$, $v_g=\prt\om(k)/\prt k$. This equation
was obtained in Ref.~\cite{el-05} in the theory of dispersive shock waves as a small-amplitude
limit of the Whitham modulation equations and in Refs.~\cite{kamch-20} from the Hamilton theory
of wave packets propagation. The general equations (\ref{eq4}) were considered in
Refs.~\cite{sk-23,kamch-23} with applications to the problems of propagation of linear wave
packets and asymptotic theory of solitons produced from an intensive initial pulse. Here we
shall consider the relation of the asymptotic integrability condition to the complete
integrability of nonlinear wave equations.

\section{Transition to the complete integrability} 

As was noticed in Ref.~\cite{ks-24},
the solution of Eqs.~(\ref{eq4}) is related with the quasiclassical limit of Lax pairs of
nonlinear wave equations completely integrable in framework of the Ablowitz-Kaup-Newell-Segur
(AKNS) scheme \cite{akns-74} written in a scalar form \cite{kk-02}. Let the completely
integrable equation can be written as a compatibility condition of two linear equations
\begin{equation}\label{eq7}
  \phi_{xx}=\mathcal{A}\phi,\qquad \phi_t=-\frac12  \mathcal{B}_x\phi+ \mathcal{B}\phi_x,
\end{equation}
where the functions $\mathcal{A}$ and $\mathcal{B}$ depend on the wave variables $\rho,u$, their
space derivatives of different orders, and on the spectral parameter $\la$. If we turn to the
quasiclassical limit of these equations with fast oscillations of $\phi$, so that we can
neglect all the derivatives of the smooth functions $\rho$ and $u$ in the expressions for
${\mathcal{A}}$ and ${\mathcal{B}}$, then we arrive at their quasiclassical limits
$\overline{\mathcal{A}}$ and $\overline{\mathcal{B}}$ which only depend on the wave variables
$\rho$ and $u$ and the spectral parameter.

The nonlinear equations under consideration reduce in the dispersionless limit to some
hydrodynamic-like equations which can be transformed to the form (\ref{eq2.1}) for the
Riemann invariants $r_+,r_-$. On the other hand, linearization of our completely integrable
equations yields the dispersion relation $\om=\om(k,r_+,r_-)$. Thus, we arrive at the
system (\ref{eq4}) which in case of completely integrable equations has a solution
\begin{equation}\label{eq8}
  k=k(r_+,r_-,q),
\end{equation}
where $q$ is an integration constant. As was shown in Ref.~\cite{ks-24},
the functions $\overline{\mathcal{A}}$ and $\overline{\mathcal{B}}$ are related with $k$
and $\om$ by the formulas
\begin{equation}\label{eq9}
\begin{split}
  \overline{\mathcal{A}}&=-\frac14k^2(r_+,r_-,q),\\
  \overline{\mathcal{B}}&=-\frac{\om(k(r_+,r_-,q),r_+,r_-)}{k(r_+,r_-,q)},
  \end{split}
\end{equation}
where $q$ is related with the spectral parameter $\la$ of the IST method. It is natural
to ask, whether it is possible to go in the opposite direction and starting from the
hydrodynamic equations and dispersion relation which satisfy the asymptotic integrability
condition (\ref{eq5}), so that the solution of Eqs.~(\ref{eq4}) is known, to arrive at
the Lax pair (\ref{eq7}) of some completely integrable nonlinear wave equation.

Actually, this procedure of generalization of quasiclassical formulas to their fully
dispersive counterparts is analogous to the well-known quantization procedure of classical
mechanical systems resulting in equations for their quantum-mechanical behavior. As is
well known, such a quantization procedure is not trivial and unique, so we shall confine
ourselves to the simplest methods widely used in physics.

In the first method, we can consider the functions $\overline{\mathcal{A}},\overline{\mathcal{B}}$
given by Eqs.~(\ref{eq9}) as exact functions defining the Lax pair (\ref{eq7}), so their
compatibility condition $(\phi_{xx})_t=(\phi_t)_{xx}$, i.e.,
\begin{equation}\label{eq10}
  \mathcal{A}_t-2\mathcal{B}_x \mathcal{A}- \mathcal{B}\mathcal{A}_x+\frac12\mathcal{B}_{xxx}=0
\end{equation}
yields the corresponding nonlinear wave system.

In the second method, we can transform Eqs.~(\ref{eq9}) to the variables $\rho,u$ and combine these
variables to a complex field
\begin{equation}\label{eq10a}
  \psi=\sqrt{\rho}\exp\left(i\int^xu(x',t)dx'\right)
\end{equation}
according to the well-known Madelung transformation \cite{madelung} of the quantum-mechanical
wave function. Then in the quasiclassical
limit we have $u\approx-i\psi_x/\psi$ or $u\approx i\psi^*_x/\psi^*$, so we can try to return
to the nonlinear dispersive equation for the field $\psi$ by the replacements
\begin{equation}\label{eq11}
  \rho\to\psi\psi^*,\quad u\to-\frac{i\psi_x}{\psi}\quad\text{or}\quad
  u\to\frac{i\psi^*_x}{\psi^*}
\end{equation}
in the quasiclassical expressions $\overline{\mathcal{A}}$ and $\overline{\mathcal{B}}$. Besides
that, we might have to add to these functions some terms vanishing in the quasiclassical limit.
The form of these term can be guessed by means of analogy with known results and correctness of
their choice can be proved by checking the compatibility condition (\ref{eq10}).

Let us illustrate this approach by several examples.

\section{Shallow water and Bose-Einstein condensate} 

Let the system be described in the
dispersionless limit by the shallow water equations
\begin{equation}\label{eq12}
  \rho_t+(\rho u)_x=0,\quad u_t+uu_x+\rho_x=0.
\end{equation}
As a dispersion relation we choose the formula
\begin{equation}\label{eq13}
  \om=k\left(u\pm\sqrt{\rho+\frac{k^2}{4}}\right)
\end{equation}
which in the long wavelength limit gives Eqs.~(\ref{eq2.2}) with $c=\sqrt{\rho}$,
that is
\begin{equation}\label{eq13b}
  v_{\pm}=u\pm\sqrt{\rho}.
\end{equation}
The Riemann invariants are defined by the formulas 
\begin{equation}\label{eq13b}
  r_{\pm}=u/2\pm\sqrt{\rho}.
\end{equation}
It is convenient to rewrite Eqs.~(\ref{eq4})
in terms of $\rho$ and $u$, so we get the equations
\begin{equation}\label{eq14}
  \frac{\prt k^2}{\prt\rho}=-4,\quad  \frac{\prt k^2}{\prt u}=-2\sqrt{k^2+4\rho},
\end{equation}
which define the function $k=k(\rho,u)$. The asymptotic integrability condition (\ref{eq5})
is fulfilled and Eqs.~(\ref{eq14}) have the solution
\begin{equation}\label{eq15}
\begin{split}
  k^2&=(q-u)^2-4\rho\\
  &=(q-2r_+)(q-2r_-).
  \end{split}
\end{equation}
If we assume that $q>u$ and choose the upper sign in Eq.~(\ref{eq13}), then we obtain
\begin{equation}\label{eq16}
  \frac{\om}{k}=\frac12(q+u).
\end{equation}

The first method of quantization yields the Lax pair (\ref{eq7}) with
\begin{equation}\label{eq17}
  \mathcal{A}=\rho-\frac14(q-u)^2,\quad \mathcal{B}=-\frac12(q+u),
\end{equation}
where $q$ plays the role of the spectral parameter. This is the Lax pair for the
Kaup-Boussinesq system \cite{kaup-75}
\begin{equation}\label{eq18}
  \rho_t+(\rho u)_x-\frac14\rho_{xxx}=0,\quad u_t+uu_x+\rho_x=0.
\end{equation}
It finds applications to dynamics of shallow water dispersive waves.

If we treat the system (\ref{eq12}) as Landau equations \cite{landau-41} for the superfluid
component and assume that Eq.~(\ref{eq13}) is the Bogoliubov dispersion relation
\cite{bogol-47} for linear waves propagating along a uniform condensate, then the second
method of quantization seems appropriate for the condensate's wave function
$\psi=\sqrt{\rho}\exp\left(i\int^xu(x',t)dx'\right)$. We make replacements (\ref{eq11}) in
Eqs.~(\ref{eq17}) and after adding a higher order term for fulfillment of Eq.~(\ref{eq10})
we arrive at the Lax pair with
\begin{equation}\label{eq19}
  \begin{split}
  \mathcal{A}&=-\frac14\left(q+\frac{i\psi_x}{\psi}\right)^2+\psi\psi^*-
  \left(\frac{\psi_x}{2\psi}\right)_x,\\
  \mathcal{B}&=-\frac12\left(q-\frac{i\psi_x}{\psi}\right).
  \end{split}
\end{equation}
This is the well-known scalar form of the Lax pair for the
Gross-Pitaevskii \cite{gross-61,pit-61} (or nonlinear Schr\"{o}dinger (NLS)
\cite{bn-67}) equation
\begin{equation}\label{eq20}
  i\psi_t+\frac12\psi_{xx}-|\psi|^2\psi=0.
\end{equation}

\section{Derivative NLS equations} 

As we saw above, the shallow water equations (\ref{eq12})
and the dispersion relation (\ref{eq13}) lead to different completely integrable systems
depending on the method of `quantization' of the solution (\ref{eq15}) of the asymptotic
integrability equations. Consequently, if we have some completely integrable equation for
a complex field $\psi=\sqrt{\rho}\exp\left(i\int^xu(x',t)dx'\right)$, then the corresponding
dispersionless system and dispersion relation for linear waves are also known and they satisfy
the asymptotic integrability condition, so the solution (\ref{eq8}) can be found together with
the quasiclassical limit (\ref{eq9}) of the functions ${\mathcal{A}}$ and ${\mathcal{B}}$.
Then they define another completely integrable system for the variables $\rho$ and $u$ which
can be considered as a `shallow water' counterpart of the equation for the complex field $\psi$. 
Here we shall consider such systems obtained from three forms of the derivative NLS (DNLS) equation.

\subsection{Kaup-Newell DNLS equation} 

The Kaup-Newell DNLS equation\cite{kn-78} has the form
\begin{equation}\label{eq21}
  i\psi_t+\frac12\psi_{xx}-i(|\psi|^2\psi)_x=0.
\end{equation}
The substitution $\psi=\sqrt{\rho}\exp\left(i\int^xu(x',t)dx'\right)$ yields in the dispersionless
limit the hydrodynamic system
\begin{equation}\label{eq22}
  \begin{split}
  & \rho_t+\left[\rho\left(u-\frac32\rho\right)\right]_x=0,\\
  &u_t+uu_x-(\rho u)_x=0.
  \end{split}
\end{equation}
The characteristic velocities are given by the formulas
\begin{equation}\label{eq23}
  v_{\pm}=u-2\rho\pm\sqrt{\rho(\rho-u)}
\end{equation}
and Eqs.~(\ref{eq22}) can be cast to the diagonal form (\ref{eq2.1}) for the Riemann invariants
\begin{equation}\label{eq24}
  r_{\pm}=\frac{u}{2}-\rho\pm\sqrt{\rho(\rho-u)}.
\end{equation}
Linearization of Eq.~(\ref{eq21}) for small-amplitude waves propagating along a uniform background
yields the dispersion relation
\begin{equation}\label{eq25}
  \om=k\left(u-2\rho\pm\sqrt{\rho(\rho-u)+\frac{k^2}{4}}\right).
\end{equation}
Eqs.~(\ref{eq4}) can be transformed to the variables $\rho,u$, so they read
\begin{equation}\label{eq26}
  \begin{split}
  &\frac{\prt k^2}{\prt\rho}=4\left[\sqrt{k^2+4\rho(\rho-u)}-(2\rho-u)\right],\\
  &\frac{\prt k^2}{\prt u}=-2\left[\sqrt{k^2+4\rho(\rho-u)}-2\rho\right]
  \end{split}
\end{equation}
and have the solution
\begin{equation}\label{eq27}
\begin{split}
  k^2 &= (q-u)^2+4q\rho\\
  &=(q-2r_+)(q-2r_-),
  \end{split}
\end{equation}
where $q$ is an integration constant, so that
\begin{equation}\label{eq28}
  \frac{\om}{k}=\frac{q}{2}+\frac{u}{2}-\rho.
\end{equation}
We define the Lax pair according to Eqs.~(\ref{eq9}),
\begin{equation}\label{eq29}
  \mathcal{A}=-\frac14(q-u)^2-q\rho,\qquad \mathcal{B}=-\frac{q}{2}-\frac{u}{2}+\rho,
\end{equation}
where $q$ plays the role of the spectral parameter. The compatibility condition yields
the completely integrable system
\begin{equation}\label{eq30}
  \begin{split}
  & \rho_t+\left[\rho\left(u-\frac32\rho\right)\right]_x+\frac{1}{4u}(u-2\rho)_{xxx}=0,\\
  &u_t+uu_x-(\rho u)_x+\frac{1}{2u}(u-2\rho)_{xxx}=0.
  \end{split}
\end{equation}
Obviously, this is a dispersive generalization of the hydrodynamic system (\ref{eq22})
different from the Kaup-Newell equation transformed to the $(\rho,u)$-variables:
\begin{equation}\label{eq31}
  \begin{split}
  & \rho_t+\left[\rho\left(u-\frac32\rho\right)\right]_x=0,\\
  &u_t+uu_x-(\rho u)_x+\left(\frac{\rho_x^2}{8\rho^2}-\frac{\rho_{xx}}{4\rho}\right)_x=0.
  \end{split}
\end{equation}
Although the substitutions
\begin{equation}\label{eq32}
  \overline{\rho}=\rho(\rho-u),\quad \overline{u}=u-2\rho
\end{equation}
transform Eqs.~(\ref{eq30}) to the Kaup-Boussinesq system (\ref{eq18}) for the variables
$(\overline{\rho},\overline{u})$, the inverse transform is not single-valued, the system
(\ref{eq30}) is not genuinely nonlinear and describes nonlinear waves different from those
for the Kaup-Boussinesq system.

\subsection{Chen-Lee-Liu DNLS equation}

The Chen-Lee-Liu equation \cite{cll-79}
\begin{equation}\label{eq33}
  i\psi_t+\frac12\psi_{xx}-i|\psi|^2\psi_x=0
\end{equation}
reduces after Madelung transformation (\ref{eq10a}) to the dispersionless equations
\begin{equation}\label{eq34}
  \rho_t+\left[\rho\left(u-\frac{\rho}{2}\right)\right]_x=0,\quad u_t+uu_x-(\rho u)_x=0
\end{equation}
and leads to the dispersion relation
\begin{equation}\label{eq35}
  \om=k\left(u-\rho\pm\sqrt{-\rho u+\frac{k^2}{4}}\right),
\end{equation}
so we get the characteristic velocities
\begin{equation}\label{eq35a}
  v_{\pm}=u-\rho\pm\sqrt{-\rho u}.
\end{equation}
The Riemann invariants can be defined by the formulas
\begin{equation}\label{eq35b}
  r_{\pm}=\frac12(u-\rho)\pm\sqrt{-u\rho}.
\end{equation}
The asymptotic integrability equations
\begin{equation}\label{eq36}
  \begin{split}
  &\frac{\prt k^2}{\prt\rho}=4u+2\sqrt{k^2-4\rho u},\\
  &\frac{\prt k^2}{\prt u}=4\rho-2\sqrt{k^2-4\rho u}
  \end{split}
\end{equation}
have the solution
\begin{equation}\label{eq37}
\begin{split}
  k^2&=(q-u+\rho)^2+4\rho u\\
  &=(q-2r_+)(q-2r_-),
  \end{split}
\end{equation}
so
\begin{equation}\label{eq38}
  \frac{\om}{k}=\frac12(q+u-\rho).
\end{equation}
As a result we arrive at the Lax pair with
\begin{equation}\label{eq39}
  \mathcal{A}=-\frac14(q-u+\rho)^2-\rho u,\qquad \mathcal{B}=-\frac12(q+u-\rho),
\end{equation}
and, consequently, obtain the completely integrable system
\begin{equation}\label{eq40}
\begin{split}
  &\rho_t+\left[\rho\left(u-\frac{\rho}{2}\right)\right]_x+\frac{1}{4(\rho+u)}(u-\rho)_{xxx}=0,\\
  &u_t+uu_x-(\rho u)_x+\frac{1}{4(\rho+u)}(u-\rho)_{xxx}=0.
  \end{split}
\end{equation}
Again this system reduces to the Kaup-Boussinesq system for the variables
\begin{equation}\label{eq41}
  \overline{\rho}=-\rho u,\quad \overline{u}=u-\rho.
\end{equation}

\subsection{Gerdjikov-Ivanov DNLS equation}

The Gerdjikov-Ivanov equation \cite{gi-83}
\begin{equation}\label{eq42}
  i\psi_t+\psi_{xx}+i\psi^2\psi^*_x+\frac12|\psi|^4\psi=0
\end{equation}
yields the dispersionless equations
\begin{equation}\label{eq43}
\begin{split}
  &\rho_t+\left[\rho\left(2u+\frac{\rho}{2}\right)\right]_x=0,\\
  &u_t+2uu_x-(\rho u)_x-\rho\rho_x=0,
  \end{split}
\end{equation}
the dispersion relation
\begin{equation}\label{eq44}
  \om=k\left(2u \pm\sqrt{-\rho(\rho+2u)+{k^2}}\right),
\end{equation}
and the characteristic velocities
\begin{equation}\label{eq44a}
  v_{\pm}=2u \pm\sqrt{-\rho(\rho+2u)}.
\end{equation}
The Riemann invariants are defined by the formulas
\begin{equation}\label{eq44b}
  r_{\pm}=u\pm\sqrt{-\rho(\rho+2u)}.
\end{equation}
The asymptotic integrability equations
\begin{equation}\label{eq45}
  \begin{split}
  &\frac{\prt k^2}{\prt\rho}=2(u+\rho),\\
  &\frac{\prt k^2}{\prt u}=2\rho-2\sqrt{k^2-4\rho(\rho+2u)}
  \end{split}
\end{equation}
give the solution
\begin{equation}\label{eq46}
\begin{split}
  k^2&=(q-u)^2+\rho(\rho+2u)\\
  &=(q-r_+)(q-r_-),
  \end{split}
\end{equation}
so
\begin{equation}\label{eq47}
  \frac{\om}{k}=q+u.
\end{equation}
Consequently, we obtain the Lax pair with
\begin{equation}\label{eq48}
  \mathcal{A}=-\frac14(q-u)^2-\frac14\rho(\rho+2u),\qquad \mathcal{B}=-q-u,
\end{equation}
and the corresponding completely integrable system
\begin{equation}\label{eq49}
\begin{split}
  &\rho_t+\left[\rho\left(2u+\frac{\rho}{2}\right)\right]_x+\frac{1}{\rho+u}u_{xxx}=0,\\
  &u_t+2uu_x-(\rho u)_x-\rho\rho_x=0.
  \end{split}
\end{equation}
which can be transformed to the Kaup-Boussinesq system for the variables
\begin{equation}\label{eq50}
  \overline{\rho}=-\rho(\rho+2u),\quad \overline{u}=2u.
\end{equation}
As is known, the above three DNLS equations are related with each other by gauge
transformations \cite{ws-83,kundu-84,kundu-87}. As we see, their `shallow water'
counterparts are related with each other by simple algebraic formulas.  At the same
time, the physical meaning of these equations can be different, as it will be 
especially clearly seen in one more example considered in the next section.

\section{Landau-Lifshitz equation and two-layer `shallow water' model} 

The Landau-Lifshitz equation
\begin{equation}\label{eq51}
  \mathbf{M}_t=(\mathbf{M}_{xx}-M_3\mathbf{e}_3)\wedge\mathbf{M}
\end{equation}
for the magnetization vector $\mathbf{M}=(M_1,M_2,M_3)$, $|\mathbf{M}|=1$, with an easy-plane
anisotropy, can be transformed to more convenient variables $w,v$ by a Madelung-like
substitution \cite{ikcp-17,ckp-16,ish-17}
\begin{equation}\label{eq52}
\begin{split}
  M_3&=-w,\\
  M_-&\equiv M_1-iM_2\\
  &=\sqrt{1-w^2}\exp\left(-i\int^xvdx'\right).
  \end{split}
\end{equation}
The resulting equations
\begin{equation}\label{eq53}
  \begin{split}
   &w_-\left[v(1-w^2)\right]_x=0,\\
   &v_t-\left[w(1-v^2)\right]_x
  +\left[\frac{1}{\sqrt{1-w^2}}\left(\frac{w_x}{\sqrt{1-w^2}}\right)_x\right]_x=0
  \end{split}
\end{equation}
reduce to the dispersionless equations
\begin{equation}\label{eq54}
  \begin{split}
   w_t&-\left[v(1-w^2)\right]_x=0,\\
   v_t&-\left[w(1-v^2)\right]_x=0
   \end{split}
\end{equation}
which can be transformed to the form (\ref{eq2.1}) by introduction of the Riemann
invariants
\begin{equation}\label{eq55}
  r_{\pm}=v w\pm\sqrt{(1-v^2)(1-w^2)}.
\end{equation}
The characteristic velocities of Eqs.~(\ref{eq54}) are equal to
\begin{equation}\label{eq56}
  v_{\pm}=2v w\pm\sqrt{(1-v^2)(1-w^2)}.
\end{equation}
The whole system (\ref{eq53}) after linearization with respect to small-amplitude
waves propagating along a uniform background yields the dispersion relation
\begin{equation}\label{eq57}
  \om=k\left(2wv\pm\sqrt{k^2+(1-v^2)(1-w^2)}\right)
\end{equation}
for harmonic waves. The asymptotic integrability equations (\ref{eq4}) written
in terms of the $(v,w)$-variables 
\begin{equation}\label{eq58}
  \begin{split}
  & \frac{\prt k^2}{\prt w}=2\left[w(1-v^2)-v\sqrt{k^2+(1-v^2)(1-w^2)}\right],\\
  & \frac{\prt k^2}{\prt v}=2\left[v(1-w^2)-w\sqrt{k^2+(1-v^2)(1-w^2)}\right],
  \end{split}
\end{equation}
have the solution
\begin{equation}\label{eq59}
\begin{split}
  k^2&=(q-v w)^2-(1-v^2)(1-w^2)\\
  &=(q-r_+)(q-r_-)
  \end{split}
\end{equation}
so (assuming $q-vw>0$) we get
\begin{equation}\label{eq60}
  \frac{\om}k=q+vw.
\end{equation}
We define the Lax pair (\ref{eq7}) with
\begin{equation}\label{eq61}
  \begin{split}
  &\mathcal{A}=\frac14[(1-v^2)(1-w^2)-(q-v w)^2],\\
  &\mathcal{B}=-q-vw
  \end{split}
\end{equation}
and obtain the corresponding completely integrable equations
\begin{equation}\label{eq62}
  \begin{split}
   w_t&-\left[v(1-w^2)\right]_x+\frac{v(vw)_{xxx}}{v^2-w^2}=0,\\
   v_t&-\left[w(1-v^2)\right]_x-\frac{w(vw)_{xxx}}{v^2-w^2}=0.
   \end{split}
\end{equation}
This is a `shallow water' counterpart of the Landau-Lifshitz equation (\ref{eq51}).

It is remarkable that the dispersionless equations (\ref{eq54}) were derived in Ref.~\cite{ovs-79}
as a long wave shallow water approximation for dynamics of inner waves in the fluid
composed of two thin layers with very small difference of densities of liquids in them (the
lighter layer above the heavier one). The dispersion relation for linear inner waves in a
two-layer system is described by a more complicated formula than Eq.~(\ref{eq57}) (see, e.g.,
Ref.~\cite{lamb}). Nevertheless, it has similar general structure, so Eqs.~(\ref{eq62})
can be considered as a `toy model' for dispersive dynamics of shallow inner waves.
Although Eqs.~(\ref{eq62}) can be reduced to the Kaup-Boussinesq system for the variables
\begin{equation}\label{eq63}
  \overline{\rho}=(1-v^2)(1-w^2),\quad \overline{u}=2vw,
\end{equation}
the qualitative properties of these two systems are very different. For example, the
Kaup-Boussinesq system is genuinely nonlinear and an initial discontinuity evolves according
to it into one of six possible wave structures \cite{eggk-95,cikp-17}. The system (\ref{eq62})
is not genuinely nonlinear and classification of possible wave structures includes various
combined dispersive shock waves similar to the structures appearing in the theory of the
Landau-Lifshitz equation \cite{ikcp-17}. One might expect that such a classification of
possible wave structures evolving from initial discontinuities in the theory of 
Eqs.~(\ref{eq62}) is qualitatively similar to such a classification for the actual two-layer
dispersion relation.

As one can see, although the considered above equations belong to different physical situations, 
they all correspond to the same solution of the asymptotic integrability equations with the
dispersion relation
\begin{equation}\label{k1}
  \om=k\left(r_++r_-\pm\sqrt{\frac14(r_+-r_-)^2+k^2}\right),
\end{equation}
so that the long wavelength characteristic velocities are given by 
\begin{equation}\label{k2}
  v_{\pm}=r_++r_-\pm\frac12(r_+-r_-),
\end{equation}
and the solution can be written in the form
\begin{equation}\label{k3}
  k^2=\left[q-\frac12(r_++r_-)\right]^2-\frac14(r_+-r_-)^2.
\end{equation}
Only the formulas connecting the Riemann invariants $r_{\pm}$ with the physical variables 
depend on the choice of the physical system and the above list of equations is not exhaustive. 
In particular, one can easily check that the
Jaulent-Miodek system \cite{jm-76,zhou-97} is described by the same solution of the
asymptotic integrability equations.

\section{Zakharov-Ito system}

To demonstrate another type of solution of the asymptotic integrability equations, let us 
consider the hydrodynamic-like equations written in the Riemann form (\ref{eq2.1}), where
\begin{equation}\label{eq65}
  v_+=6r_++2r_-,\quad v_-=2r_++6r_-.
\end{equation}
One can check that the asymptotic integrability condition is fulfilled if we choose
the dispersion relation
\begin{equation}\label{eq66}
\begin{split}
  \om&=k\Big\{4(r_++r_-)+{k^2}/{2}\\
  &\pm\sqrt{\left[2(r_++r_-)+{k^2}/2\right]^2-16r_+r_-}\Big\},
  \end{split}
\end{equation}
so that Eqs.~(\ref{eq4}) have the solution
\begin{equation}\label{eq67}
  k^2=\frac{4}{q}(q-r_+)(q-r_-),
\end{equation}
and then
\begin{equation}\label{eq68}
  \frac{\om}{k}=4q+2(r_++r_-).
\end{equation}
Let the Riemann invariants be related with the `physical' variables $\rho,u$ by the formulas
\begin{equation}\label{eq69}
  r_+=\frac12(u+\sqrt{u^2-4\rho}),\quad r_-=\frac12(u-\sqrt{u^2-4\rho}),
\end{equation}
so that
\begin{equation}\label{eq70}
  k^2=4\left(q-u+\frac{\rho}q\right),\quad \frac{\om}{k}=4q+2u.
\end{equation}
Transition to the Lax pairs according to the rules (\ref{eq9}) yields
\begin{equation}\label{eq71}
  \mathcal{A}=-q+u-\frac{\rho}{q},\quad \mathcal{B}=-4q-2u,
\end{equation}
where $q$ plays the role of the spectral parameter. The resulting completely integrable
system is the well-known Zakharov-Ito system \cite{zakh-80,ito-82}
\begin{equation}\label{eq72}
  \begin{split}
  & \rho_t+2u\rho_x+4\rho u_x=0,\\
  & u_t+6uu_x-4\rho_x-u_{xxx}=0.
  \end{split}
\end{equation}
It is worth noticing that although the dispersionless system with the
velocities (\ref{eq65}) written in $(\rho,u)$-variables
\begin{equation}\label{eq73}
  \begin{split}
  & \rho_t+2u\rho_x+4\rho u_x=0,\\
  & u_t+6uu_x-4\rho_x=0
  \end{split}
\end{equation}
can be transformed to the shallow water system (\ref{eq12}) for the variables
\begin{equation}\label{eq74}
  \overline{\rho}=4u^2-16\rho,\quad \overline{u}=4u,
\end{equation}
these substitutions are not compatible with transformation of the dispersion relation
(\ref{eq13}). In fact, they cast the dispersion relation (\ref{eq66}) to the form
\begin{equation}\label{eq75}
  \om=k\left(4u+\frac{k^2}{2}\pm\sqrt{\left(2u+\frac{k^2}{2}\right)^2-16\rho}\right)
\end{equation}
different from the relation (\ref{eq13}) for the Kaup-Boussinesq system.

\section{Conclusion}

In this paper, we introduced the physically natural condition of asymptotic integrability of
nonlinear wave equations which means that two asymptotic limits---dispersionless
(hydrodynamic) evolution of a smooth background wave and propagation of high-frequency
wave packets---are compatible with each other in the sense that the Hamiltonian structure
of the packet's propagation is preserved by evolution of the background wave and, consequently, 
the Hamilton equations have an integral of motion. In fact,
this weaker integrability condition leads to the quasiclassical limit of complete
integrability in framework of the AKNS scheme \cite{ks-24} and if the Lax pair of the
equations under consideration does not depend on the space derivatives of the wave variables,
then the asymptotic integrability condition reproduces the exact Lax pair. This observation
sheds new light on the origin of the completely integrable nonlinear wave equations.

Besides that, it should be noticed that the relationship for the dependence of the packet's
wave number on the background variables can be converted to a similar relationship for the
inverse half-width of narrow solitons propagating along smooth background, and this remark
allows one to develop a Hamiltonian theory for propagation of narrow solitons along a
non-uniform and time dependent background \cite{ks-23,kamch-24b,ks-24b}.

At last, Eqs.~(\ref{eq4}) of asymptotic integrability can also have approximate solutions
correct in the limit of large wave numbers $k$ even for not completely integrable
equations. In these quite general situations, one can develop the theory of propagation of
small-amplitude wave packets along a non-uniform and time-dependent background \cite{sk-23}
and to formulate a generalized Bohr-Sommerfeld quantization rule which determines the 
parameters of solitons produced from an intensive initial pulse \cite{kamch-23}. Thus, the
condition of asymptotic integrability turns out to be a useful tool for investigation of
various problems in nonlinear physics.

\begin{acknowledgments}

I thank E.~A.~Kuznetsov and M.~V.~Pavlov for useful discussions.
This research is funded by the research project FFUU-2021-0003 of the Institute of Spectroscopy
of the Russian Academy of Sciences (Sections 1-3) and by the RSF grant number~19-72-30028
(Sections~4-7).

\end{acknowledgments}

\section*{Data Availability Statement}

The data that support the findings of this study are available
from the author upon reasonable request.

\end{document}